\documentstyle{article}
\begin{document}

\author{Leonid B. Litinskii}
\title{Some rigorous results for the Hopfield Model}
\date{High Pressure Physics Institute of Russian Academy of 
Sciences\\
Russia, 142190 Troitsk Moscow region, e-mail: 
litin@hppi.troitsk.ru}
\maketitle
\begin{abstract}
\it For the Hopfield model with the Hebb connection matrix we investigate the 
case of $p$ memorized patterns that are distorted copies of the same {\it 
standard}. In other words, we try to simulate that learning always takes place 
by means of repeating presentations of one and the same standard, and the 
presentations are accompanied by distortions of the standard. 
We obtain some rigorous results relating to the dependence of the fixed points 
on external parameters of the problem. 
\end{abstract}

\section{Introduction}
By the Hopfield model of a neural network we mean a dynamic system of $n$ 
interacting "agents". The agents can take only two values $\sigma_i=\pm 1,\
i=1,2,\ldots,n$, and they are connected with each other by the Hebb connection 
matrix  
$${\bf J=S^{\rm T}\cdot S},\ J_{ii}=0,\ i=1,2,\ldots,n.\eqno(1)$$
The matrix $\bf S$ is $(p\times n)$-matrix whose $n$-dimensional rows are the 
memorized patterns used for the network learning.

Let us consider the following meaningful problem: a network had to be taught by
$p$-time presentations of {\it the standard}, but errors crept into the learning
process and, in fact, the network was taught by $p$ distorted copies of the 
standard. What is the distortion's influence on the result of the learning?

We formalize the problem choosing the matrix $\bf S$ in the form,
$$\bf S =\left(\begin{array}{ccccccc}
1-x&1&\ldots&1&1&\ldots&1\\
1&1-x&\ldots&1&1&\ldots&1\\
\vdots&\vdots&\ddots&\vdots&\vdots&\ldots&\vdots\\
1&1&\ldots&1-x&1&\ldots&1\end{array}\right).
\eqno(2)$$
The $n$-dimensional rows of the matrix $\bf S$,
$$\vec s^{(l)}=(1,\ldots,\underbrace{1-x}_l,1,\ldots
,1)\in {\rm R}^n,\ l=1,\ldots,p\eqno(3)$$
are treated as $p$ distorted copies of {\it the standard}
$$\vec\varepsilon (n)=(1,\ldots,1)\in {\rm R}^n.\eqno(4)$$ 
The real number $x$ is called {\it the distortion parameter}. 
 
Our goal is to find out how the set of fixed points depends on the 
distortions of the standard (because the set of fixed points represents the
"memory" of the network). In other words, we investigate 
the ability of the Hopfield model to generalization \cite{Fon} in the specific
case of one distorted standard.    

Only the first $p$ coordinates of the standard are distorted, and the remained 
$q=n-p$ coordinates are not distorted. The problem (1)-(4) is discussed in 
Section 2. Of course, the distortions are rather specific: for each memorized 
pattern only one coordinate of the standard is distorted and the value of the 
distortion $x$ is the same for all the memorized patterns. In the next Sections
we generalize this simple model.

The asynchronous dynamics of the network is supposed. The state of the network 
as a whole is described by {\it the configuration vector} 
$$\vec\sigma=(\sigma_1,\sigma_2,\ldots,\sigma_n).$$

\section{Basic Model}

The problem (1)-(4) will be called {\it the Basic Model}. It was investigated
in detail in \cite{Lit}. The results of \cite{Lit} are summarized in items
A)-C). 

{\bf A).} {\it A necessary condition for a configuration vector to be a fixed 
point is that its last $q$ coordinates be equal to each other.}

Consequently, fixed points can be chosen in the form
$$\vec\sigma^*=(\underbrace{\sigma_1,\sigma_2,\ldots,\sigma_p}_
{\vec\sigma^\prime},1,\ldots,1);\eqno(5)$$
Here, $\vec\sigma^\prime=(\sigma_1,\sigma_2,\ldots,\sigma_p)$ is the
$p$-dimensional part of the configuration vector $\vec\sigma^*$. This
$p$-dimensional vector will be used below.

{\bf B).} {\it If a vector $\vec\sigma^*$ (5) is a fixed point, then every 
configuration vector of the form (5) whose distance from the standard is the 
same as the distance of $\vec\sigma^*$ is a fixed point.} 

Consequently, we have to divide all the set of $2^p$ vectors $\vec\sigma^*$
into {\it classes} $\Sigma$ joining vectors equidistant from the standard. 
For the standard $\vec\varepsilon (n)$ these classes are
$$\Sigma_k=\left\{\vec\sigma^*\left| \sum_{i=l}^p\sigma_l=p-
2k\right.\right\},\ 
k=0,1,\ldots,p.\eqno(6)$$  
The number of the classes $\Sigma_k$ is equal to $p+1$. The number of the 
vectors in the class $\Sigma_k$ is equal to ${p\choose k}$. The configuration 
vectors from the same class $\Sigma_k$ are the fixed points simultaneously. 

\noindent
{\bf C).} \underline{\bf Theorem 1:}

\noindent
{\it
When $x$ increases from $-\infty$ to $\infty$, in consecutive order the set of 
the fixed points is exhausted by the classes  
$$\Sigma_0,\Sigma_1,\ldots,\Sigma_{k_{max}}.$$
The $k$th rebuilding from the class $\Sigma_{k-1}$ to the class $\Sigma_k$ 
occurs at the point $x_k$:
$$x_k=p\cdot\frac{n-(2k-1)}{n+p-2(2k-1)},\quad 
k=1,2,\ldots,k_{max},\eqno(7)$$
where}
$$k_{max}= \min\left(p,\left[\frac{n+p+2}4\right]\right)=
\left\{\begin{array}{cl}p,&\mbox{ 
when }\frac{p-1}{n-1}<\frac13;\\
\left[\frac{n+p+2}4\right],&\mbox{ when }\frac{p-1}{n-1}>\frac13.
\end{array}\right.\eqno(8)$$
In Fig.1 the graphical illustration of Theorem 1 is shown.

The case $p=n$ worth to be mentioned specially. Here all the rebuilding points
$x_k$ stick to one point,
$$x_k\equiv\frac{n}2,\ k=1,2,\ldots,\left[\frac{n+1}2\right].\eqno(9)$$
For any $x$ from the left side of $\frac{n}2$ there is only one fixed point. It
is the standard $\vec\varepsilon (n) = \Sigma_0$. For $x$ from the right side
of $\frac{n}2$ every configuration vector from the class 
$\Sigma_{\left[\frac{n+1}2\right]}$ is the fixed point; in other words, in this
region there are ${n\choose \left[\frac{n+1}2\right]}$ fixed points.

\section{Complete Generalization}

$\bf 1^\circ.$ The standard $\vec\varepsilon (n)$ can be replaced by an 
arbitrary configuration vector  
$$\vec{\alpha}=(\alpha_1,\alpha_2,\ldots,\alpha_n),\ \alpha_i=\{\pm 1\}.$$
In this case the memorized patterns have the form
$$\vec s^{(l)}=(\alpha_1,\ldots,(1-x)\alpha_l,\ldots,\alpha_n),\ 
l=1,2,\ldots,p.$$
All the statements of items A)-C) remain valid, but now the vectors
$\vec\sigma^*$ (5) have the form
$$\vec\sigma^*=(\alpha_1\sigma_1,\alpha_2\sigma_2,\ldots,\alpha_p\sigma_p,
\alpha_{p+1},\ldots,\alpha_n).$$
Thus, $\vec\varepsilon (n)$ can be used as the standard without the loss of 
generality.

$\bf 2^\circ.$ Suppose the memorized patterns are obtained from the standard
$\vec\varepsilon (n)$ by {\it identical and simultaneous} distortions of its
$m$ (but not only one!) coordinates,
$$\vec s^{(l)}=(\underbrace{1,\ldots,1}_m,\ldots,
\underbrace{1-x,\ldots,1-x}_
{\begin{array}{l}\mbox{\footnotesize $l$th group of $m$}\\ 
\mbox{\footnotesize coordinates}\end{array}},
\ldots,\underbrace{1,\ldots,1}_m,\underbrace{1,\ldots,1}_q),\ l=1,2,\ldots,p.$$ 
Of course, now $n=p\cdot m+q$. It is not difficult to show that the fixed
points  
have the form of the piecewise constant vectors
$$\vec\sigma^*=(\underbrace{\sigma_1,\ldots,\sigma_1}_m,\underbrace{
\sigma_2,\ldots,\sigma_2}_m,\ldots,
\underbrace{\sigma_p,\ldots,\sigma_p}_m,\underbrace{1,\ldots,1}_q).$$
As above these vectors are grouped into classes $\Sigma_k^{(m)}$ analogous to 
the classes $\Sigma_k$ (6). The only difference is that now the value $-1$ has 
to be assigned not to a separate coordinate $\sigma_l$, but to $m$ coordinates 
of the same name. Again, the number of the vectors $\vec\sigma^*$ in the class 
$\Sigma_k^{(m)}$ is equal to $p\choose k$. Then we have the generalization of 
the Theorem 1:

\noindent
{\it The value of the parameter $x$ corresponding to the $k$th rebuilding from 
the class $\Sigma_{k-1}^{(m)}$ to the class $\Sigma_k^{(m)}$ is} 
$$x_k=p\cdot\frac{\frac{n}m-(2k-1)}{\frac{n}m+p-2(2k-1)},\
k=1,2,\ldots,k_{max},\  
k_{max}= \min\left(p,\left[\frac{\frac{n}m+p+2}4\right]\right).$$

$\bf 3^\circ.$ The memorized patterns (3) can be rotated as a whole. 
Suppose the rotation matrix ${\bf U}=(u_{ij})$ acts on the first $p$
coordinates  
of $n$-dimensional vectors only. Then the standard takes the form
$$\vec u=(\underbrace{u_1,u_2,\ldots,u_p}_{\vec u^\prime},1,\ldots,1),
\eqno(10)$$
where
$$u_l=\sum_{i=1}^p u_{li},\ l=1,2,\ldots,p.\eqno(11)$$ 
In Eq.(10) the vector $\vec u^\prime$ is the $p$-dimensional part of new
standard $\vec u$ (the same as the vector $\vec\sigma^\prime$ in 
Eq.(5) is the $p$-dimensional part of the vector $\vec\sigma^*$). It is easy to
see that $\parallel\vec u^\prime\parallel^2=p$.

At the same time, the memorized patterns (3) take the form
$$\vec s^{(l)}=(u_1-xu_{1l},u_2-xu_{2l},\ldots,
u_p-xu_{pl},1,\ldots,1),\ l=1,2,\ldots,p.\eqno(12)$$
The elements of the relevant connection matrix ${\bf J}^{(U)}$ are
$$J^{(U)}_{ij}=J_{ij}a_ia_j,\mbox{ where
}a_i=\left\{\begin{array}{cl} u_i&,\mbox{ when }1\le i\le p\\
1&,\mbox{ when }p< i\le n,\end{array}\right.$$
and the matrix elements $J_{ij}$ are determined by Eqs.(1),(2).

Let us suppose that the standard $\vec\varepsilon (n)$ remains unchanged after 
the rotation: $$u_l\equiv 1.$$ 
Because the connection matrix ${\bf J}^{(U)}$ coincides with the initial 
matrix $\bf J$, all the 
statements of the items A)-C) of the Basic Model remain unchanged too. Note, 
here all the first $p$ coordinates of the memorized patterns (12) are 
distorted. 
Consequently, we succeeded in getting rid of one of the limitation of the Basic
Model (see Introduction).

\section{Ground State}

It is well-known \cite{Her}, that {\it an energy} 
$E=-\frac1n\sum_{i,j=1}^nJ_{ij}\cdot\sigma_i\cdot\sigma_j$ can be associated 
with every state $\vec\sigma$ of a network. The energy decreases during the 
network evolution. The fixed points are the local minima of the energy $E$. A 
fixed point that is the global minimum of the energy is called {\it the ground 
state} of a network. 

It is not difficult to show that for the cases discussed above, all the 
states that are the fixed points simultaneously have the same energies. In
other words, every time the set of fixed points is the ground state of the
network, and there are no local extremums. This is due to the symmetry of the
problems in  
question. When the symmetry of the problem is reduced, it is possible that 
different fixed points have different energies. Often just the 
ground state is of interest, because its energy is minimal. From my point of 
view namely the ground state has to be treated as the result of learning.
Only the ground state is discussed in this Section. 

$\bf 1^\circ.$ Let us examine the problem of the item $3^\circ$ from the 
previous Section for the case when the standard $\vec\varepsilon (n)$ changes 
due to rotation: 
$$u_l\not\equiv 1.$$ 
Again, as in the item A), only configuration vectors $\vec\sigma^*$ (5) can be 
the fixed points. Again, the statement of the item B) is valid, however now it 
is true not for all the fixed points, but for the ground state of the network 
only:

\noindent
{\it If a vector $\vec\sigma^*$ (5) is a ground state, then every configuration
vector of the form (5) whose distance from the standard $\vec u$ 
is the same as the 
distance of $\vec\sigma^*$ is a ground state.} 

Consequently, we have to divide the set of the vectors $\vec\sigma^*$ into 
classes joining the vectors that are equidistant from the standard $\vec u$.
Only now 
the vectors that are equidistant from $\vec u$ are vectors 
$\vec\sigma^*$ with the same value of the cosine of the angle between 
$p$-dimensional vectors $\vec\sigma^\prime$ and $\vec u^\prime$,
$$\cos  w=\frac{\sum_{i=1}^p \sigma_i\cdot u_i}p=
\frac{(\vec\sigma^\prime,\vec u^\prime)}{\parallel\vec\sigma^\prime 
\parallel\cdot\parallel\vec u^\prime\parallel}.\eqno(13)$$
Let the vectors $\vec\sigma^*$ be grouped into classes $\Sigma_k^{(U)}$,
$$\Sigma_k^{(U)}=\left\{\vec\sigma^*\left| \sum_{i=1}^p \sigma_i\cdot u_i=
p\cdot\cos w_k\right.\right\},\ k=0,1,\ldots,t,$$
where $t+1$ is the number of the different values of the cosine (13). 
Without loss of generality we can assume that the cosines are arranged in 
decreasing order,
$$\cos w_0>\cos w_1>\ldots>\cos w_t.$$ 
Then it is easy to see, that $\cos w_k=-\cos w_{t-k},\ \forall k\le t$.
Therefore, the cosines are negative beginning from some number $k$.
Finally, the generalization of the Theorem 1 from the item C) is: 

\noindent
{\it When $x$ increases from $-\infty$ to $\infty$, in consecutive order the 
ground state is exhausted by the classes  
$$\Sigma^{(U)}_0,\Sigma^{(U)}_1,\ldots,\Sigma^{(U)}_{k_{max}}.$$
The $k$th rebuilding of the ground state from the class $\Sigma^{(U)}_{k-1}$ to
the class $\Sigma^{(U)}_k$ occurs at the point $x_k$, 
$$x_k=\frac{p}{2}\left[1+\frac{q}{q+p\cdot(\cos w_{k-1}+\cos w_k)}\right],
\ k=1,2,\ldots,k_{max}.\eqno(14)$$ 
If $x_1>\frac34p$, then $k_{max}=t$. If $x_1<\frac34p$, the rebuildings come to 
an end when the denominator in Eq.(14) becomes negative ($k_{max}<t$).}

\underline{\bf Note.} When the distortion $x$ belongs to the interval 
$(x_k,x_{k+1})$, all the configuration vectors from the class $\Sigma^{(U)}_k$ 
are the ground state of the network. But the compositions of the classes 
$\Sigma^{(U)}_k$ are determined by the values of $\{u_l\}_{l=1}^p$ (11) only. 
And the choice of $\{u_l\}_{l=1}^p$ is in the researcher's hand. In other 
words, 
selecting $\{u_l\}_{l=1}^p$ and the distortion parameter $x$, the network with
a preassigned ground state can be created.

For example, let us create a network with the ground state 
$$\begin{array}{rrrrrrrrl}\vec\alpha=(&-1,&-1,&1,&1,&1,&\ldots,&1&)\\
\vec\beta=(&1,&1,&-1,&1,&1,&\ldots,&1&)\\
\vec\gamma=(&1,&1,&1,&-1,&1,&\ldots,&1&).\end{array}$$
At first we need to find the vector $\vec u$ of the form (10) equidistant from 
the vectors $\vec\alpha, \vec\beta$ and $\vec\gamma$ (and only from these 
vectors!). This problem has a lot of solutions. For example, we can take the 
vector $\vec u$ in the form of
$$\vec u=(u,u,2u,2u,u_5,\ldots,u_p,1,\ldots,1),\mbox{ where }
10u^2+\sum_{l=5}^pu_l^2=p.$$
The minimal $p$ satisfying this equation is equal to $4$. Consequently, we have
$$\vec u=(u,u,2u,2u,\underbrace{1,\ldots,1}_q),\ u=\sqrt\frac25,\ q\ge 0.$$
Next, $2^4$ configuration vectors $\vec\sigma^*$ (5) split into classes 
$\Sigma^{(U)}_k$, which join the vectors equidistant from the found standard 
$\vec u$. One of these classes consists from our vectors $\vec\alpha,
\vec\beta$  
and $\vec\gamma$ exactly. It is not difficult to determine the number of this 
class. Indeed, the first four values of the cosines (13) and the relevant 
classes $\Sigma_k^{(U)}$ are:
$$\begin{array}{l}
\begin{array}{l}\cos w_0=6u/p=3u/2\\
\begin{array}{rrrrrl}\Sigma_0^{(U)}=(&1,&1,&1,&1,&1,\ldots,1)\in{\rm{\bf R}^
{4+q}}
\end{array}\end{array}\\
\\
\begin{array}{l}\cos w_1=\cos w_0-2u/p=u\\
\Sigma_1^{(U)}=\left\{\begin{array}{rrrrrl}(&-1,&1,&1,&1,&1,\ldots,1)\\
(&1,&-1,&1,&1,&1,\ldots,1)\end{array}\right.\end{array}\\
\\
\begin{array}{l}\cos w_2=\cos w_0-4u/p=u/2\\
\Sigma_2^{(U)}=\left\{\begin{array}{rrrrrl}
(&-1,&-1,&1,&1,&1,\ldots,1)\\
(&1,&1,&-1,&1,&1,\ldots,1)\\
(&1,&1,&1,&-1,&1,\ldots,1)\end{array}\right.\end{array}\\
\\
\begin{array}{l}\cos w_3=\cos w_0-6u/p=0\\
\Sigma_3^{(U)}=\left\{\begin{array}{rrrrrl}
(&-1,&1,&-1,&1,&1,\ldots,1)\\
(&-1,&1,&1,&-1,&1,\ldots,1)\\
(&1,&-1,&-1,&1,&1,\ldots,1)\\
(&1,&-1,&1,&-1,&1,\ldots,1)\end{array}\right.\end{array}
\end{array}$$
Thus, the class $\Sigma_2^{(U)}$ contains the given configuration vectors 
$\vec\alpha$, $\vec\beta$ and $\vec\gamma$ only. If we take four memorized 
patterns of the form (12) and the distortion parameter $x$ inside the interval 
$(x_2,x_3)$, the relevant network has the preassigned ground state.  

To define the memorized patterns $\vec s^{(l)}$, we need the matrix elements 
$u_{kl}$ of the rotation matrix $\bf U$. This matrix transforms the vector 
$\vec\varepsilon (n)$ into the standard $\vec u$. More exactly, it is necessary
to know only the upper $(4\times 4)$-block of this matrix realizing the 
nontrivial part of the rotation. In other words, it is necessary to solve the 
problem
$${\bf U}^{(4\times 4)}=\left(\begin{array}{cccc}u_{11}&u_{12}&u_{13}&u_{14}\\
u_{21}&u_{22}&u_{23}&u_{24}\\
u_{31}&u_{32}&u_{33}&u_{34}\\
u_{41}&u_{42}&u_{43}&u_{44}\end{array}\right)\cdot
\left(\begin{array}{c}1\\1\\1\\1\end{array}\right)=
\left(\begin{array}{c}u\\u\\2u\\2u\end{array}\right),\mbox{ where }
u=\sqrt{\frac25}.$$
This problem has a lot of solutions. For example,
$${\bf U}^{(4\times 4)}=\left(\begin{array}{cccc}-u/2&0&3u/2&0\\0&3u/2&0&-u/2\\
3u/2&0&u/2&0\\0&u/2&0&3u/2\end{array}\right).$$
Consequently, as the memorized patterns (12) $(4+q)$-dimensional vectors 
$$\begin{array}{rccccl}\vec s^{(1)}=(&u(1+x/2),&u,&u(2-3x/2),&2u,&1,\ldots,1)\\
\vec s^{(2)}=(&u,&u(1-3x/2),&2u,&u(2-x/2),&1,\ldots,1)\\
\vec s^{(3)}=(&u(1-3x/2),&u,&u(2-x/2),&2u,&1,\ldots,1)\\
\vec s^{(4)}=(&u,&u(1+x/2),&2u,&u(2-3x/2),&1,\ldots,1),\end{array}$$
can be used. If $x$ is from the interval 
$2\left(1+q/(q+6u)\right)<x<2\left(1+q/(q+2u)\right)$,
the configuration vectors $\vec\alpha$, $\vec\beta$ and $\vec\gamma$ are the 
ground state of the Hopfield network with the connection matrix 
(1). The minimal $n$ for such a network is equal to $5$, $n=5$. This
corresponds to $q=1$. If $q=0$, we have the equality $p=n$, and, consequently, 
all the 
rebuilding points stick to the point $\frac{n}2$ (see the comment to Eq.(9)). 

$\bf 2^\circ.$  Suppose the asynchronous dynamics is defined as
$$\sigma_i(t+1)={\rm sgn}\left(\frac1{n}\cdot\sum_{j=1}^n 
J_{ij}\sigma_j(t)+H\right),$$ 
where the connection matrix $\bf J$ is given by Eqs.(1),(2).  
The parameter $H$ is called {\it the dynamic threshold} \cite{Her}. Usually, in
theoretical considerations it is used to eliminate the linear term in the
energy $E$ (such a term appears when we turn from the $(0,1)$-network to the
$(-1,+1)$-network). Apparently, the role of the dynamic threshold is much more
important.  
In \cite{Lit2} the case of $H \ne 0$ is analyzed in details. For the sake of 
simplicity, here I present the results obtained for $H>0$ only.

\noindent
\underline{\bf Theorem 2.} 

\noindent
{\it Let $x\in(x_k,x_{k+1})$. When $H$ increases from its initial zero value to 
infinity, in consecutive order the ground state of a network is exhausted by
the classes
$$\Sigma_k,\ \Sigma_{k-1},\ \Sigma_{k-2},\ldots,\ \Sigma_0.$$
The rebuilding of the ground state from the class $\Sigma_{k+1-i}$ to the class
$\Sigma_{k-i}$ occurs when $H$ is equal to}
$$H_{k+1-i}(x)=\frac{n+p-2(2(k+1-i)-1)}{n}\cdot(x-x_{k+1-i}),\ 
i=1,2,\ldots,k.$$
Here, the classes $\Sigma_k$ are defined by Eq.(6) and the rebuilding points
$x_k$ are given by Eq.(7). In Fig.2 for some values of $p$ and $n$ the phase 
diagram for the ground state is shown. We see that changing $H$ purposefully,
we can change the ground state significantly. In particular, the standard 
$\vec\varepsilon (n)=\Sigma_0$ can be done the ground state of the network. It 
is important to understand either this result depends on the specific form of 
our connection matrix, or it is more general. In other words, is it possible to
choose the thresholds in the general Hopfield model in such a way that
memorized  
patterns will necessarily be its fixed points? The positive answer seems to be 
very probable. 

$\bf 3^\circ.$ I generalize the Basic Model for the case of different 
distortions of all the memorized patterns \cite{Lit1}:
$$\vec s^{(l)}=(1,\ldots,1,\underbrace{1-x^{(l)}}_l,1,\ldots,1),\ 
l=1,2,\ldots,p.$$
Suppose, that
$$x^{(1)} > x^{(2)}>\ldots > x^{(p_+)}>0>x^{(p_++1)}>\ldots
> x^{(p)},$$
and $p_+$ is the number of positive distortions $x^{(l)}$. 
Theorem 3 is an analog of the statements A)-C) from the Basic Model.

\noindent
\underline{\bf Theorem 3.}

\noindent
{\bf a)} {\it Only configuration vectors
$$\vec\sigma^*(k)=(\underbrace{-1,-1,\ldots,-1}_{k},1,\ldots,1),\ 
k=0,1,\ldots,k_{max},\eqno(15)$$ 
can be the ground state, where}
$$k_{max}= \min\left(p_+,\left[\frac{n+p+2}4\right]\right).$$

\noindent
{\bf b)} {\it For the vector $\vec\sigma^*(k)$ to be a ground state, it is 
necessary and sufficient to have}
$$x^{(k+1)}-\frac{\sum\limits^k_{i=1} x^{(i)}-\sum\limits^p_{j=k+2} x^{(j)}}
{n+p-2(2k+1)}<\ p <\ x^{(k)}-\frac{\sum\limits^{k-1}_{i=1} x^{(i)}-
\sum\limits^p_{j=k+1}x^{(j)}}{n+p-2(2k-1)}.\eqno(16)$$

The inequalities (16) are very useful. For example, suppose $p\ll n$, and $n$
is so large that in Eq.(16) the terms containing $n$ in the denominators can
be omitted. Then, the statement b) is simplified:
$$\vec\sigma^*(k)\ \mbox{\it is the ground state if and only if }\ 
x^{(k+1)}<\ p\ <x^{(k)}.\eqno(17)$$
In particular, 
$$\vec\sigma^*(0)\ \mbox{\it is the ground state if and only if }\ 
x^{(1)}< p.\eqno(18)$$
These statements are very interesting. Namely, Eqs.(17) and (18) connect the 
quality of learning with the relation between the number of presentations $p$
of the standard and the values of distortions. (The distortions are measured in
bits.) In particular, Eq.(18) means that the network understands the standard 
$\vec\varepsilon (n)=\vec\sigma^*(0)$ correctly, if and only if the number of 
its presentations $p$ exceeds the maximum distortion $x^{(1)}$. It seems, this 
result clarifies qualitatively why the well-known Latin saying {\it "Repetitio 
est mater studiorum"} is true. Indeed, to be sure that a network with the Hebb 
connection matrix understands a standard, we have to show it repeatedly. Also, 
we must be sure that the number of presentations $p$ is greater than the
maximal distortion $x^{(1)}$. Since usually the distortions are Gaussianly
distributed, the last requirement can be realized always.

Next, we can rewrite the inequalities (16) in the form 
$$x_k < x^{(k)}-d^{(k)};\ x^{(k+1)}-d^{(k+1)}<x_{k+1},\eqno(19)$$
where
$$d^{(k)}=\frac{\sum_{i=1}^p|x^{(i)}-x^{(k)}|}{n+p-2(2k-1)}.$$
Sometimes Eqs.(19) are more suitable, since they relate the distortions 
$x^{(k)}$ with the rebuilding points $x_k$ from Eq.(7). When the distortions
are known, with the aid of Eqs.(19) the ground state can be easily found 
graphically. 

Let us clarify the last statement with the aid of a simple
example. Let $p=5$ and $n$ is so large that $k_{max}=5$ and $d^{(k)}\propto
o(1/n)$. Only one of the configuration vectors
$\vec\sigma^*(k),\ k=0,1,\ldots,5$ of the form  
(15) can be the ground state. We make use of Eq.(19). We write down the 
necessary and sufficient conditions for each $\vec\sigma^*(k)$ to be a ground 
state (here $d^{(k)}$ can be omitted):
$$\begin{array}{ccl}\vec\sigma^*(0)&\Longleftrightarrow&x^{(1)}<x_1.\\
\vec\sigma^*(1)&\Longleftrightarrow&x_1<x^{(1)};\ x^{(2)}<x_2.\\
\vec\sigma^*(2)&\Longleftrightarrow&x_2<x^{(2)};\ x^{(3)}<x_3.\\
\vec\sigma^*(3)&\Longleftrightarrow&x_3<x^{(3)};\ x^{(4)}<x_4.\\
\vec\sigma^*(4)&\Longleftrightarrow&x_4<x^{(4)};\ x^{(5)}<x_5.\\
\vec\sigma^*(5)&\Longleftrightarrow&x_5<x^{(5)}.\end{array}$$
Next, let the rebuilding points $x_k$ and the distortions $x^{(k)}$ be located 
as it is shown in Fig.3. Since only for $k=2$ the inequalities (19) are 
fulfilled, the vector $\vec\sigma^*(2)=(-1,-1,1,\ldots,1)$ is the ground state.

\section{Summary}

We present rigorous results relating to the model (1)-(4) 
and its generalizations. Computer simulations confirm these results. The proofs
of the statements of Sections 2-4 are in \cite{Lit}, \cite{Lit2}, \cite{Lit1}.
A meaningful interpretation of the obtained results can be found in the same 
references. In the present publication we do not include this topic
deliberately with the exception of "the proof" of the Latin saying.  

The work was supported by Russian Foundation for Basic 
Research (grant 01-01-00090).

The author is grateful to Prof. Alexandr Ezhov for helpful discussions
and to Dr. Inna Kaganova for preparation of the
manuscript. Wonderful atmosphere and personal contacts with the members of
workshop "Statistical Physics of Neural Networks" 
(Max-Plank-Institute of Physics of Complex Systems, March 1999, Dresden,
Germany) helped the author in the understanding of the obtained results.

\vskip 2cm
\centerline{\bf List of Figures}
\vskip 1cm
\noindent
{\bf Fig.1.} The rebuilding points $x_k$ from Eq.(7) divide the axis $x$ into
intervals inside which the set of the fixed points belongs to different classes
$\Sigma_k$ (see Theorem 1). 
\vskip 5mm
\noindent
{\bf Fig.2.} The straight lines $H_i(x)$ from Theorem 2 divide the half-plane
$(x,H>0)$ into regions inside which the ground state belongs to different
classes $\Sigma_i$; $p=3,\ n=9$.
\vskip 5mm
\noindent
{\bf Fig.3.} The example of the location of the rebuilding points $x_k$
from Eq.(7) and the different distortions $x^{(k)}$ from Theorem 3.


\newpage
\begin{thebibliography}{99}
\bibitem{Fon} J.F.Fontanari. "Generalization in a Hopfield network", {\it
Journal de Physique (France)}, 1990, v.51, pp.2421-2430. 
\bibitem{Lit} L.B.Litinskii. "High-symmetry Hopfield-type neural networks", 
{\it Theoretical and Mathematical Physics}, Kluwer Academic/Plenum 
Publishers, 
1999, v.118, pp. 107-127. Cond-mat/9906197. 
\bibitem{Her} J.Hertz, A.Krogh, R.Palmer. {\it Introduction to the Theory of 
Neural Computation}, Addison-Wesley, 1991.
\bibitem{Lit2} L.B.Litinskii. "Hopfield model with threshold", 
{\it Theoretical and Mathematical Physics} (2001, in press).
\bibitem{Lit1} L.B.Litinskii. (in preparation).
\end{thebibliography}
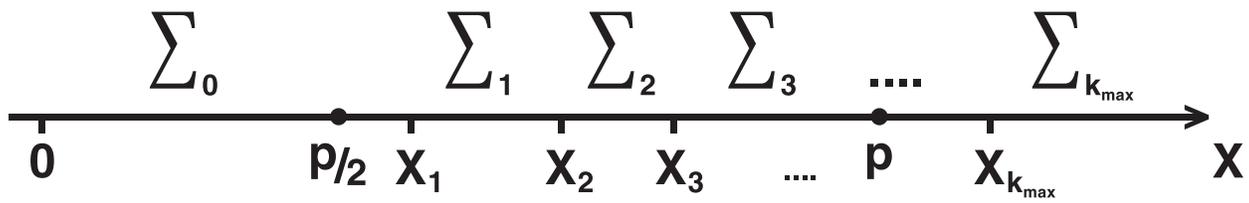
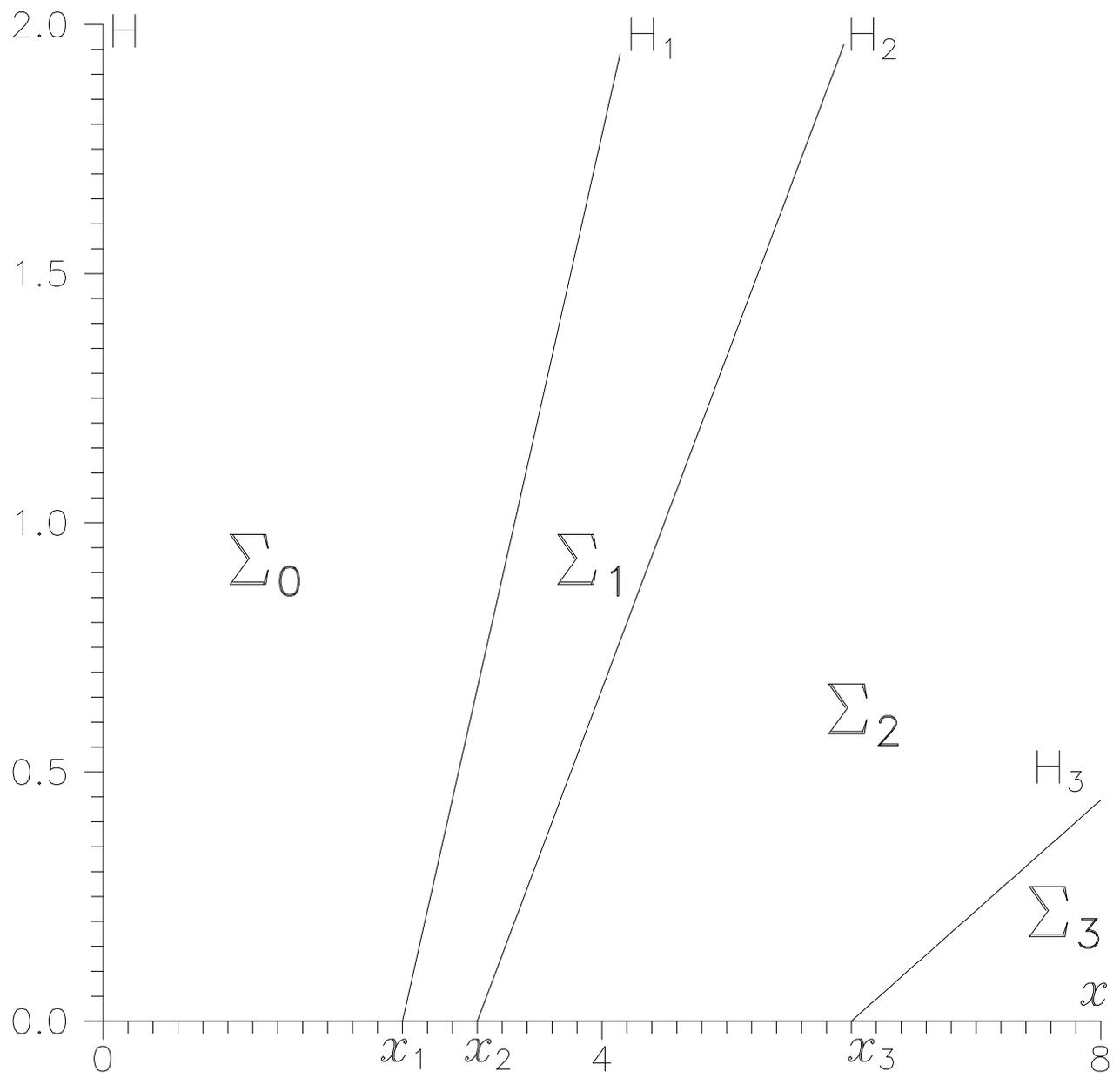
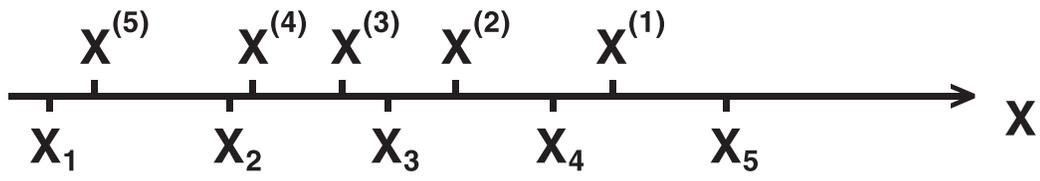
\end{document}